# Structural changes in Ti$_{1-x}$Al$_x$N coatings during turning: A XANES and EXAFS study of worn tools

L. Rogström[1*], M. Moreno[1], J. Andersson[2], M. Johansson-Jõesaar[2], M. Odén[1], K. Klementiev[3], L.-Å. Näslund[4], M. Magnuson[4]

[1]*Nanostructured Materials, Dept. Physics, Chemistry and Biology (IFM), Linköping University, 581 83 Linköping, Sweden*

[2]*Seco Tools AB, 737 82 Fagersta, Sweden*

[3]*MAX IV Laboratory, Lund University, SE-221 00 Lund, Sweden*

[4]*Thin Film Physics, Dept. Physics, Chemistry and Biology (IFM), Linköping University, 581 83 Linköping, Sweden*

* Corresponding author: lina.rogstrom@liu.se

**Abstract**

Structural changes in Ti$_{1-x}$Al$_x$N coated tool inserts used for turning in 316L stainless steel were investigated by XANES, EXAFS, EDS, and STEM. For coarse-grained fcc-structured Ti$_{1-x}$Al$_x$N coatings, with 0≤x≤0.62, the XANES spectrum changes with Al-content. XANES Ti 1*s* line-scans across the rake face of the worn samples reveals that TiN-enriched domains have formed during turning in Ti$_{0.47}$Al$_{0.53}$N and Ti$_{0.38}$Al$_{0.62}$N samples as a result of spinodal decomposition. The XANES spectra reveal the locations on the tool in which the most TiN-rich domains have formed, indicating which part of the tool-chip contact area that experienced the highest temperature during turning. Changes in the pre-edge features in the XANES spectra reveal that structural changes occur also in the w-TiAlN phase in fine-grained Ti$_{0.38}$Al$_{0.62}$N during turning. EDS shows that Cr and Fe from the steel adhere to the tool rake face during machining. Cr 1*s* and Fe 1*s* XANES show that Cr is oxidized in the end of the contact length while the adhered Fe retains in the same fcc-structure as that of the 316L stainless steel.

1. Introduction

Titanium aluminum nitride (TiAlN) is one of the most important coating materials for increasing the service life and performance of tool inserts in metal machining. It is well suited for high-speed cutting operations due to its favorable high temperature behavior in terms of phase stability and mechanical properties. The spinodal decomposition of the face centered cubic (fcc) TiAlN phase and subsequent formation of wurtzite (w) AlN that occurs when the material is exposed to high temperatures is well characterized, see for example Refs. [1-5]. The decomposition of the metastable cubic TiAlN phase influences the mechanical properties and thus the wear behavior of the coating [6, 7]. Decomposition of the coating takes place during machining [8, 9] and it has been observed that interdiffusion of species from the workpiece into the coating is enhanced in the decomposed structure [10, 11]. The small contact area between the tool and the workpiece material, and the large temperature and stress gradients [8], limit the understanding on how the decomposition affects the wear behavior as the characterization of the nanostructure formation requires advanced sample preparation by focused ion beam (FIB) in combination with scanning transmission electron spectroscopy (STEM). This restricts the possibilities of mapping out the material behavior across the contact zone between the tool and the workpiece material.

Non-destructive techniques such as x-ray diffraction (XRD) [2-4, 12] and x-ray absorption spectroscopy (XAS) techniques [13-18] have successfully been used to separate between single phase and two-phase materials for as-deposited or post-annealed coatings. We have previously observed that XRD is not sensitive enough to detect the initial stage of decomposition because of the small difference in lattice parameter between the as formed cubic phase AlN and TiN-rich domains [2]. X-ray absorption near-edge structure (XANES) and extended x-ray absorption fine structure (EXAFS) are, on the other hand, more sensitive to the short-range order compared to diffraction as they probe the local environment, which changes character at early stages of decomposition. Using a combination of XANES, EXAFS and XRD, Tuilier *et al.* [13] studied magnetron sputtered TiAlN coatings and observed that Ti presents a hexagonal-like local order for high Al-contents ($Ti_{0.14}Al_{0.86}N$) while for lower Al-contents ($Ti_{0.50}Al_{0.50}N$) there is a cubic-like local order. XANES of the Al 1s- and Ti 1s-edges revealed that there is a discrepancy between the structure obtained by XRD and EXAFS, understood as while XRD only probes well crystallized domains, also distorted or amorphous-like domains, e.g., grain boundaries, contribute to the EXAFS signal [14]. The sensitivity of EXAFS to non-crystalline domains was used to confirm an increased disorder within the hexagonal TiAlN domains during nanoindentation of TiAlN coatings [15]. This could not be observed by selected area electron diffraction (SAED) in a combined TEM/SAED study since only crystalline domains are probed by SAED. Gago *et al.* could identify the presence of a wurtzite phase by recording spectra at the N 1$s$, Al 1$s$ and Ti 2$p$ edges by XAS on magnetron sputtered TiAlN films [16, 18]. The segregation of TiAlN upon annealing could also be identified by XANES [17].

In the present work, we investigate high-resolution Ti 1s XANES and EXAFS data from different positions across the contact area of the rake face of worn Ti$_{1-x}$Al$_x$N coated tools, utilizing the excellent spatial resolution available at the Balder beamline at MAX IV [19]. The XANES spectra changes with the Al-content of the coating which enables us to identify small variations in the chemical composition of the fcc-TiAlN phase. We use this to show that decomposition occurs in the crater region of the worn Ti$_{1-x}$Al$_x$N coatings with highest Al-content.

## 2. Experimental details

### 2.1. Sample deposition and cutting tests

Cobalt-toughened tungsten carbide (WC-Co) inserts (TPUN 160308E30-K) were coated with Ti$_{1-x}$Al$_x$N in a Metaplas MZR323 arc deposition system. Four separate depositions using TiAl-alloy cathodes with different Ti:Al ratio were performed for the four different coatings. The depositions were performed in nitrogen gas (N$_2$) atmosphere at the pressure of 3.5 Pa, a substrate temperature of 500 °C, and using a negative substrate bias between 25 and 40 V. Two depositions were performed with 33:67 Ti:Al cathodes of different sizes (diameter of 63 and 100 mm, respectively) to obtain coarse-grained and fine-grained Ti$_{0.38}$Al$_{0.62}$N. More details on the deposition can be found in Ref. [11].

The coated inserts were used for longitudinal turning of 316L stainless steel using a cutting speed of 220 m/min, depth of cut of 2 mm, and feed of 0.2 mm/rev. Three edges were used for each sample, and they were run for 0.5, 1, and 3 min, respectively. After initial analysis by scanning electron microscopy (SEM) and energy dispersive x-ray spectroscopy (EDS), the edges used for 1 min of turning were selected for further experiments.

XANES data was also collected for magnetron sputtered (MS), single crystal reference TiN and fcc-Ti$_{0.35}$Al$_{0.65}$N coatings. The reference coatings were grown on MgO substrates ((111) orientation for TiN and (001) orientation for Ti$_{0.35}$Al$_{0.65}$N) from a Ti or Ti$_{0.35}$Al$_{0.65}$ target (99.9% purity) in an Ar/N$_2$ gas mixture. For more details on the deposition conditions, the reader is referred to Refs. [20] and [21].

### 2.2. XANES and EXAFS

The Ti 1s, Fe 1s, and Cr 1s XANES and Ti 1s EXAFS spectra were measured using a Si(111) crystal monochromator at the wiggler beamline Balder on the 3 GeV electron storage ring at MAX IV in Lund, Sweden. The data was collected in fluorescence yield mode using a 7-element SDD detector (X-PIPS, Mirion Technologies) and an 80° incidence angle on the sample. The beam size was defined by slits to 50 µm in the vertical direction. The horizontal width of the beam was approximately 200 µm. A 6 µm thick Ti metal foil was used for energy calibration where the first inflection point in the first derivative of the Ti 1s absorption spectrum was set to 4.9660 keV. The measurements were performed using scans with 0.2 eV step size and 0.1 s dwell time per step. The total intensity was accumulated over three scans for Cr 1s and Fe 1s and five scans for Ti 1s. The lower statistics for the Ti 1s XAS was a result of de-tuning the monochromator to reduce the contribution of unwanted undulator-derived harmonics. The XANES data from the seven channels was summed, background corrected by fitting a linear function to the pre-edge data, and normalized by setting the highest intensity to one. The procedure was repeated ten times at each position at the sample and a summation provided the final XAS spectrum. A metal bulk sample was measured for each of the three elements. For Fe and Cr, a 316L steel reference was measured with the same parameters. For each worn insert, a line scan was performed across the rake face as illustrated in Figure 3.

EXAFS data was measured at the Ti *1s* absorption edge with a step size of 0.25 eV for 0.1 s per step. The EXAFS analysis was made by using the Visual Processing in EXAFS Researches (VIPER) software package [22]. The Ti-N, Ti-Ti, and Ti-Al scattering paths obtained from the Effective Scattering Amplitudes (FEFF) [23, 24] were included in the EXAFS fitting. The $k^2$-weighted χ EXAFS oscillations were extracted from the raw absorption data, the average of 10 absorption spectra, after removing known monochromator-induced glitches and subsequent atomic background subtraction and normalization. The bond distances (R), number of neighbors (N), Debye-Waller factors ($\sigma^2$, representing the amount of disorder) and the reduced $\chi r^2$ as the squared area of the residual, were determined by fitting the back-Fourier-transform signal between k=0-12 Å$^{-1}$ originally obtained from the forward Fourier-transform within R=0–3.2 Å of the first coordination shell using a Hanning window function [23, 24] with a many body factor of $S_0^2$=0.8.

## 2.3. Microstructural characterization

The phase content was determined by XRD in the Bragg-Brentano as well as grazing incidence (GI) configuration using Cu Kα radiation (Bruker D8 Advance). The lattice parameters were extracted from the fcc-111 diffraction peak by fit of a pseudo-Voigt function to the θ-2θ diffractogram to extract the peak position. The corresponding Ti-N and Ti-Ti/Ti-Al interatomic distances were calculated assuming a NaCl structure. SEM and EDS were performed in a Leo 1550 Gemini (Zeiss) instrument equipped with an Oxford X-MAX detector. The SEM was operated at 10 kV for imaging and EDS. Lamellas for STEM were prepared by FIB in a Zeiss Neon 40 instrument using the lift-out technique. They were coated by a thin Pt layer to protect the samples and were coarse milled using a 30 kV/2 nA Ga$^+$ ion beam followed by final polishing using 200 pA and 50 pA currents. The TEM lamellas were viewed in a FEI Tecnai G2 TF 20 UT instrument operating at 200 kV. STEM images were recorded using the HAADF detector and a camera length of 140 mm.

## 3. Results and discussion

Figure 1 shows GIXRD diffractograms of the as-deposited coatings. The coarse-grained coatings have a cubic structure while the fine-grained (FG) coating reveals a mixture of fcc and wurtzite phases. All coatings have a columnar structure and a thickness from 2.5-2.7 µm [11]. The chemical composition specified in Figure 1 was determined by EDS. The nitrogen-to-metal ratio as determined by EDS was close to one and similar between the samples and is not considered further here. For more details on the coating microstructure and properties, the reader is referred to Ref. [11].

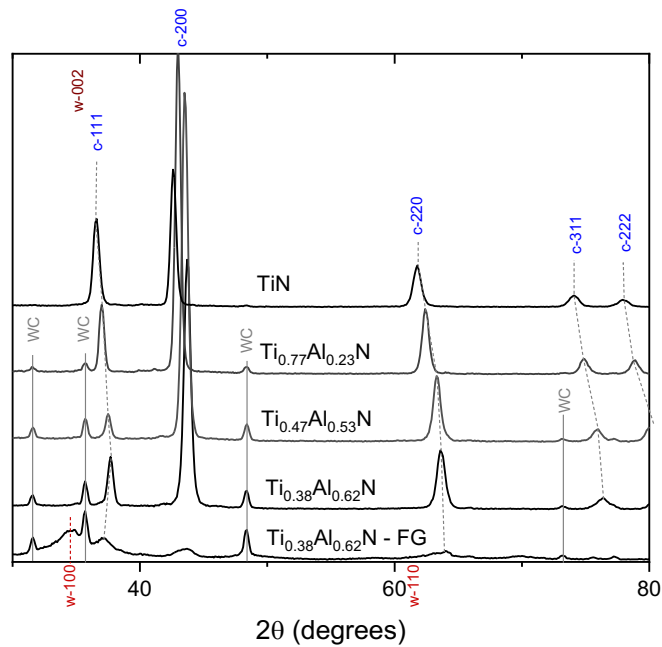

**Figure 1:** GIXRD diffractograms of the as-deposited coatings. "c-" indicates peaks from fcc-Ti(Al)N and "w-" from w-AlN.

Figure 2 shows the XANES signal for the as-deposited coatings and three reference materials. Seven different features are observed for TiN, labeled A-D and F-H in Figure 2a. The XANES signal for the arc deposited TiN coating largely resembles that of the MS-TiN reference coating (Fig. 2b). As observed, all samples show a high-energy shift of the absorption edge peak C compared to bulk Ti metal reference (Fig. 2b), indicating higher valence states as a result of charge-transfer to N and Al. The local environment of Ti atoms is expected to change in the vicinity of defects such as dislocations and grain boundaries. The higher defect-density of the arc deposited sample and the presence of grain boundaries does, however, not seem to influence the XANES spectra. Table 1 lists the energy corresponding to the intensity maximum of the peaks A, C, D, and F-H and the features appear at the same energy for both the arc-evaporated and MS reference TiN coatings. The main-edge region shows insensitivity toward the Al content. The large signal changes, observed when the Al-content increases in the arc-deposited coatings, are visual at the post-edge region. For the coarse-grained samples, increasing the Al-content above x=0.23 causes the features G and H to disappear. Endrino *et al*. observed that the features at ~5060 eV and ~5090 eV (G and H) disappear when the Al-content of the fcc-TiAlN phase decreased during annealing [17]. There is also a broadening and shift to higher energies of F, similar to what was observed by Endrino *et al*. with the change of Al-content in the fcc-TiAlN phase [17]. For the highest Al-content, x=0.62, a new feature E appears. The reference MS-$Ti_{0.35}Al_{0.65}N$ sample also show a broadening of feature F compared to TiN and a loss of feature G (Fig. 2b), thus similar to the arc deposited coatings. There is, however, a larger shift of feature F for the arc deposited films when x increases from x=0 to x=0.62, compared to when x increases from 0 to 0.65 for the MS reference coatings. Further, the feature F is broader for the arc deposited coating than for the MS reference coating. One reason for this can be the higher defect density and the presence of grain boundaries in the arc deposited coatings, resulting in a locally more disordered structure compared to the MS reference coating. However, the origin of feature F would need to be identified for a better interpretation of the exact position and broadening of this peak.

The pre-edge peak A, which actually is a double feature assigned to the Ti 1s transitions to the empty Ti-N $t_{2g}$-$e_g$ orbitals [25,26], is located at 4971 eV for all coarse-grained samples and only small changes of this feature is observed for the coarse-grained samples in the inset of Fig. 2a, indicating that the strength of the Ti-N bonds is rather insensitive toward the Al content. For $Ti_{0.47}Al_{0.53}N$ and $Ti_{0.38}Al_{0.62}N$, the intensity of the pre-edge feature has decreased and it consists of a peak at 4971 eV and a clear shoulder at 4968 eV. Thus, the $t_{2g}$-$e_g$ crystal-field splitting of the Ti 3d orbitals [25, 26] is best observed for these samples and the two shoulders suggest a $t_{2g}$-$e_g$ crystal-field splitting of about 3 eV.

The sharp pre-edge peak for the fine-grained $Ti_{1-x}Al_xN$ coating, located at 4969 eV, can be interpreted as Ti 1s transitions to unoccupied Ti 3$d$-4$p$ hybridized orbitals [27, 28] that is superposed on the $t_{2g}$-$e_g$ features. The Ti 3$d$-4$p$ hybridized orbitals feature suggests Ti-Ti or Ti-Al interaction in the fine-grained film that is absent in the coarse-grained film. One hypothesis could be that the appearance of the Ti-Ti or Ti-Al interaction is related to the increased number of grain boundaries and thus larger total interface area in the fine-grained material compared to the coarse-grained materials. However, there is no pronounced difference in the intensity of the pre-edge feature between the single-crystal MS reference samples (Fig. 2b) and the coarse-grained arc deposited samples (Fig. 2a). Thus, the grain boundaries are not contributing to the appeared Ti-Ti or Ti-Al interaction in the fine-grained coating. Furthermore, the fine-grained coating consists of a mixture of cubic and wurtzite phases that have been suggested causing dipole allowed electronic transitions into 3*d*-4*p* mixing orbitals for four-fold coordinated atoms in a tetrahedral coordination [13, 14]. Also our results presented here suggest that 3*d*-4*p* hybridized orbitals appear in the XANES spectrum because of a w-TiAlN phase while being absent for an fcc-TiAlN phase. This observation agrees with that of Endrino *et al.* [17], that also observed the sharp pre-edge peak when a wurtzite phase was present in the coating. They assigned the pre-edge peak to Ti-*d* states hybridized with Al-*p* states. It is however not possible to determine from the current data whether the 3*d* contribution originates from neighboring Al or Ti atoms.

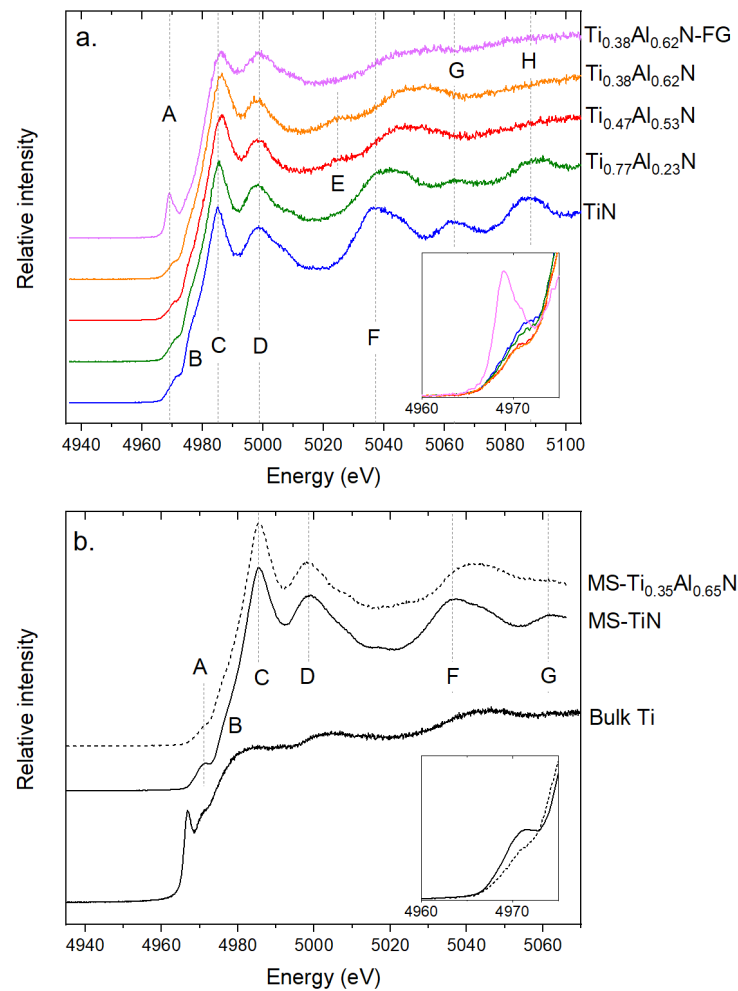

**Figure 2:** XANES spectra for the (a) as-deposited coatings and (b) reference materials. The features are labeled A-H. The insets show the pre-edge region for the TiN and TiAlN samples.

**Table 1:** Position of the intensity maximum of features A-H in the XANES spectra of the MS reference coatings, the as-deposited samples (Fig. 2), and the middle of the crater region of the worn samples (orange spectra in Fig. 5). The absorption energies are expressed in electron volt (eV).

| References | A | C | D | F | G | H |
|---|---|---|---|---|---|---|
| MS-TiN | 4971 | 4985 | 4999 | 5037 | 5062 | |
| MS- $Ti_{0.35}Al_{0.65}N$ | 4971 | 4985 | 4999 | 5043 | | |
| **As deposited** | A | C | D | F | G | H |
| TiN | 4972 | 4985 | 4998 | 5037 | 5063 | 5087 |
| $Ti_{0.77}Al_{0.23}N$ | 4972 | 4985 | 4998 | 5042 | 5063 | 5090 |
| $Ti_{0.47}Al_{0.53}N$ | 4971 | 4986 | 4998 | 5047 | | |
| $Ti_{0.38}Al_{0.62}N$ | 4971 | 4986 | 4998 | 5050 | | |
| $Ti_{0.38}Al_{0.62}N$-FG | 4969 | 4986 | 4998 | ~5048 | | |
| **Worn - crater** | A | C | D | F | G | H |
| TiN | 4972 | 4985 | 4998 | 5037 | 5062 | 5087 |
| $Ti_{0.77}Al_{0.23}N$ | 4972 | 4986 | 4998 | 5039 | 5063 | 5089 |
| $Ti_{0.47}Al_{0.53}N$ | 4972 | 4986 | 4998 | 5040 | | 5091 |
| $Ti_{0.38}Al_{0.62}N$ | 4972 | 4986 | 4998 | 5039 | 5066 | 5090 |
| $Ti_{0.38}Al_{0.62}N$-FG | 4969 | 4986 | 4999 | 5041 | | 5089 |

Figure 3 shows an SEM micrograph of the rake face of the $Ti_{0.47}Al_{0.53}N$ coating after 1 min of turning. The contact length is approximately 500 μm. The black rectangles show the measurement positions for XANES. At the position of the filled rectangle EXAFS experiments were performed. The position of each scan is illustrated in Figures 5 and 8. Figure 3b shows the corresponding EDS maps of selected elements. It can be observed that across the contact length of the tool, Fe and Cr species from the steel has adhered to the surface. At the end of the contact where the coating is in exposed to air while the temperature still is relatively high, oxide phases commonly form [11] and also in this case we observe an enrichment of O and Cr at this position. In addition to Cr, Fe and O, Ni has also adhered at the tool surface together with smaller amounts of Mg, Ca, and P (not shown).

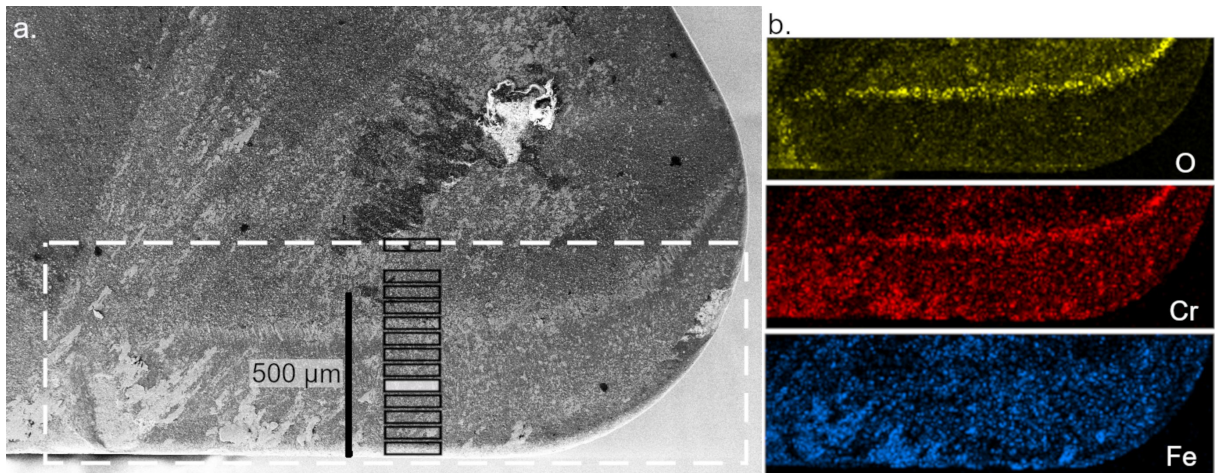

**Figure 3:** (a) SEM micrograph of the $Ti_{0.47}Al_{0.53}N$ coated insert used for 1 min of turning. The black outlined rectangles mark XAS measurement positions. (b) Elemental EDS maps from the area within the white dashed rectangle.

Figure 4 shows the magnitude of the Fourier transform from the $k^2$-weighted EXAFS oscillations $\chi(k)$ from selected as-deposited and worn samples. For the as-deposited TiN sample, the first two peaks are assigned to Ti-N and Ti-Ti bonds, respectively. Table 2 shows the results of the fitting with the interatomic distances for the Ti-N, Ti-Ti and Ti-Al scattering paths. For the two as-deposited samples of TiAlN at the bottom of Fig. 4 with x=0.53 and 0.62, the Ti-Ti/Ti-Al peak around 2.5 Å, has largely suppressed intensity as an effect of destructive interference between Ti-Ti and Ti-Al scattering paths of similar length due to the difference in their phase shift close to 180° [14]. It is also observed that the Ti-N distance becomes shorter when Al is incorporated in the fcc structure (see Table 2), as also shown by Tuilier *et al.* [14].

There is some discrepancy between the interatomic distances extracted from EXAFS and XRD. This is likely an effect of that the two methods probe short- vs long-range order, respectively, together with presence of strain in the coating. XRD in the Bragg-Brentano configuration probes only lattice planes parallel to the surface while EXAFS averages over all directions. In the presence of residual stress in the coating the lattice parameter will be altered in different directions. For the as-deposited coatings, there is a compressive in-plane strain [11], as commonly observed for arc deposited coatings. Thus, the lattice spacing is largest in the growth direction which is the direction probed by XRD here. This explains the larger interatomic distances obtained by this method. Tuilier *et al.* also reported a deviation between the lattice parameter determines by EXAFS and XRD [14].

For the worn TiAlN coatings, there is an increase of the Ti-N interatomic distance compared to the as-deposited coatings. This indicates that the structure of the coating has changed during use of the tool. As observed in Figure 4, there is a clear peak from the Ti-Ti/Ti-Al pairs around 2.5 Å. The appearance of this peak for the worn samples suggests that the Ti-Ti and Ti-Al phase contributions to the total signal has a more uneven share of Ti and Al compared to the as-deposited samples.

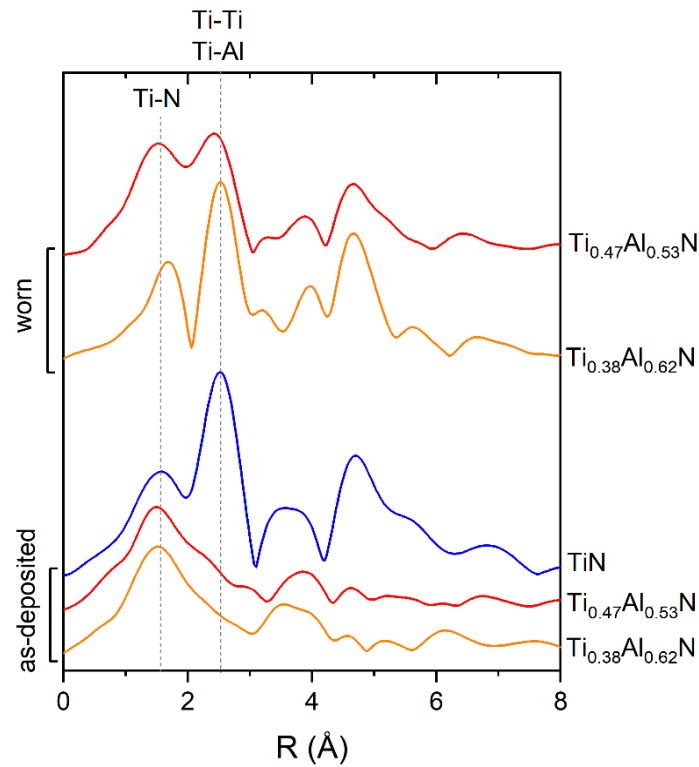

**Figure 4:** Fourier transform obtained from the $k^2$-weighted EXAFS oscillations χ($k$) of selected as-deposited and worn samples.

Figure 5 shows the Ti 1$s$ XANES data for all worn coatings as a function of distance from the edge of the tool. The color-coding in the small SEM micrograph in Figure 5f illustrates the position where each spectrum was recorded. The white arrow in Figure 5f has a length of 500 µm and illustrates the increasing distance from the tool edge, also corresponding to the chip flow direction. The center of the crater wear region is represented by the dark orange curve. This position was determined based on the abrasive wear marks observed by SEM and is the position on the rake face where the crater is the deepest. For TiN and $Ti_{0.77}Al_{0.23}N$, the spectra are very similar to those of the as-deposited coatings with the same compositions (Fig. 2). The spectra do not change across the contact, thus the fcc structure of the coatings is retained for both samples. For the higher Al-content samples, on the other hand, there is a change in the spectrum across the contact. For both $Ti_{0.47}Al_{0.53}N$ and $Ti_{0.38}Al_{0.62}N$ (Fig. 5d-e), the feature F shifts to lower energies in the middle of the contact length (orange) compared to the region closest to the edge (green) and the end of the contact (dark blue).

Feature F is found at lower energies close to the edge compared to the end of the contact and compared to the as-deposited coating. For $Ti_{0.38}Al_{0.62}N$, feature G appears in the middle of the contact in the crater region.

Figure 6 shows the position of the maximum intensity of feature F extracted from the spectra of the as-deposited and worn coatings. For the as-deposited coatings, the position of this feature is observed to shift to higher energies as more Al is introduced into the coating. This is observed both for arc-deposited and MS reference coatings (marked as stars). The exception is the fine-grained coating, for which part of the Ti is present in the w-TiAlN structure. Thus, in agreement with previous studies [17], the position of feature F is dependent on the Al content of the fcc phase. The position of feature F shifts to lower energies in the middle of the crater for all samples with Al content higher than x=0.23. Thus, this indicates that there is a Ti-rich fcc-Ti(Al)N phase forming in the crater region during turning.

**Table 2:** Interatomic distances extracted from EXAFS data from as-deposited and worn arc-deposited coatings and MS reference coatings. For the as-deposited arc coatings, the interatomic distances extracted from XRD are also presented. The two rightmost columns show data from the literature. The interatomic distances are expressed in angstrom (Å).

| | | This work – EXAFS | | | This work – XRD | Tuilier [13] | Gîrleanu [15] |
|---|---|---|---|---|---|---|---|
| **TiN** | | MS-TiN | TiN as dep. | | TiN as dep. | TiN | TiN |
| | Ti-N | 2.09±0.02 | 2.11±0.02 | | 2.13±0.04 | 2.11±0.02 | 2.13±0.02 |
| | Ti-Ti | 2.97±0.02 | 2.97±0.02 | | 3.01±0.04 | 2.99±0.02 | 3.01±0.02 |
| **$Ti_{~0.50}Al_{~0.50}N$** | | | $Ti_{0.47}Al_{0.53}N$ as dep. | $Ti_{0.47}Al_{0.53}N$ worn | $Ti_{0.47}Al_{0.53}N$ as dep. | | |
| | Ti-N | | 2.00±0.04 | 2.03±0.04 | 2.08±0.04 | | |
| | Ti-Ti | | 2.90±0.04 | 2.90±0.04 | 2.94±0.04 | | |
| | Ti-Al | | 3.00±0.04 | 3.00±0.04 | 2.94±0.04 | | |
| **$Ti_{~0.35}Al_{~0.65}N$** | | MS-$Ti_{0.35}Al_{0.65}N$ | $Ti_{0.38}Al_{0.62}N$ as dep. | $Ti_{0.38}Al_{0.62}N$ worn | $Ti_{0.38}Al_{0.62}N$ as dep. | $Ti_{0.32}Al_{0.68}N$ | $Ti_{0.32}Al_{0.68}N$ |
| | Ti-N | 2.07±0.04 | 2.01±0.04 | 2.09±0.04 | 2.07±0.04 | 2.10±0.04 | 2.08±0.03 |
| | Ti-Ti | 2.89±0.04 | 2.90±0.04 | 2.94±0.04 | 2.92±0.04 | 2.96±0.04 | 2.94±0.04 |
| | Ti-Al | 2.94±0.04 | 2.97±0.04 | 3.02±0.04 | 2.92±0.04 | 2.96±0.04 | 2.94±0.04 |

For the fine-grained sample, there is also a shift of the feature F to lower energies in the middle of the crater (Fig 5e). The inset in Figure 5e shows the pre-edge features for selected positions along the edge. In the middle to end of the contact length (orange to yellow), there is a change in shape of the pre-edge features. The intensity of the feature at 4968 eV decreases indicating less contributions from neighboring Ti or Al atoms. Based on the discussion above, where we assigned the Ti-Ti or Ti-Al interaction as specific for the w-TiAlN phase, this is interpreted as that there are changes in the w-TiAlN phase during turning. The w-TiAlN phase has been observed to have a better thermal stability compared to the fcc-TiAlN phase [29], however, there is still a lack of knowledge on the influence of temperature and pressure of the stability of this phase. For the coarse-grained samples, a wurtzite phase forming during turning is expected to be free of Ti and can thus not be detected here [29].

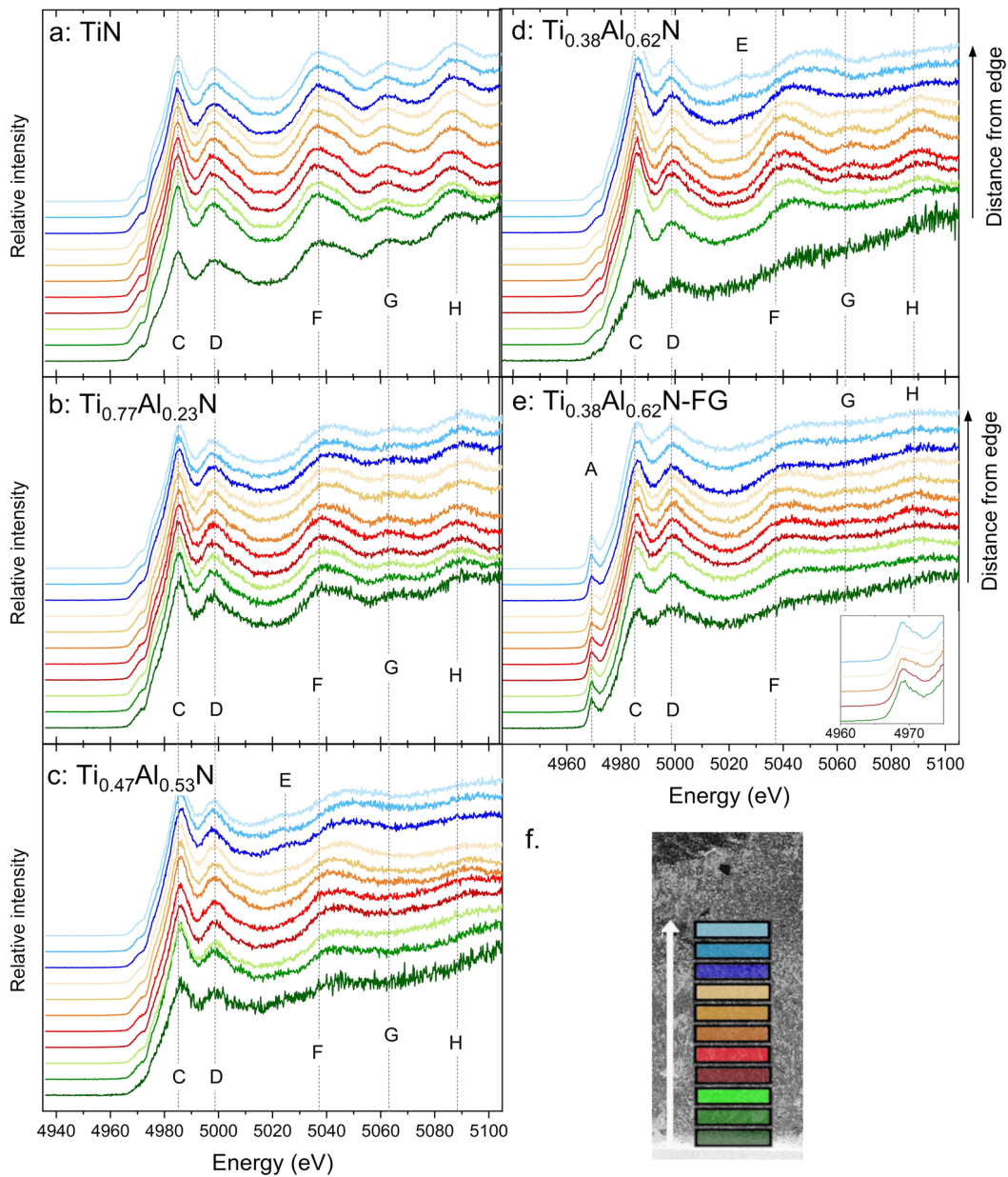

**Figure 5:** (a-e) XANES data from the Ti 1$s$-edge of the worn samples. (f) illustrates the position of the x-ray beam for each scan.

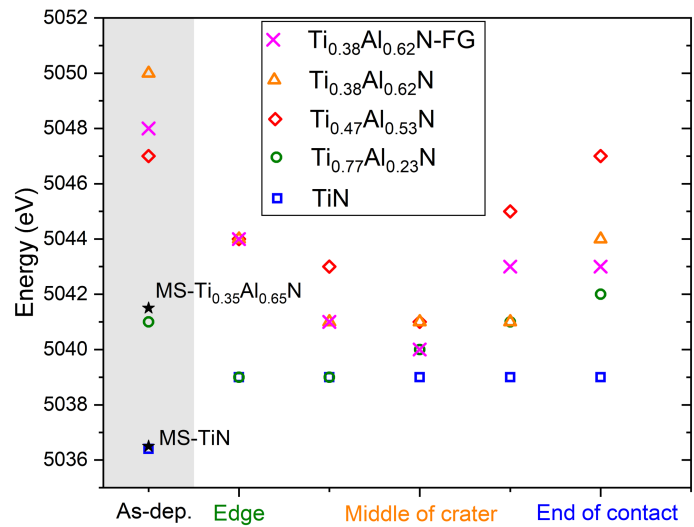

**Figure 6:** Position of feature F for the as-deposited coatings and selected positions on the worn edge. The black stars mark the values for the MS reference samples.

Figure 7 shows elemental contrast STEM micrographs from the middle of the crater (position of dark orange box in Fig. 5) of the $Ti_{0.47}Al_{0.53}N$ and $Ti_{0.38}Al_{0.62}N$ samples. For both samples, nm-sized bright and dark features are observed in the images that stem from spinodal decomposition during the turning operation causing domains with different composition. The formation of Ti-rich fcc-TiAlN domains means that a large proportion of the probed Ti-atoms are in a Ti-rich environment, thus the contribution to the XANES spectra from these Ti-atoms is the strongest. In the crater region of the worn samples, the broadness of feature F can be affected by contribution of domains with a different Ti-content. However, also for the as-deposited coatings this feature broadens with decrease of Ti in the fcc phase.

The position of feature F for the $Ti_{0.47}Al_{0.53}N$ and $Ti_{0.38}Al_{0.62}N$ coatings, has shifted to its lowest value approximately 300-350 µm from the tool edge. Interpreting the shift of this feature as an increased size of TiN-rich domains thus suggests that this is where the highest temperature is reached during use of the tool, leading to the most progressed spinodal decomposition at this position. The shift of feature F close to the edge indicates that there could be a depletion of Al from the fcc phase also at this position of the coating.

For the $Ti_{0.47}Al_{0.53}N$ coating, feature F is found at approximately the same position as that for the as-deposited sample in the end of the tool-chip contact, thus no structural changes of the coating have occurred. However, across the rest of the contact length, the shift of this feature indicates that there are domains enriched in TiN forming. For the highest Al-content, $Ti_{0.38}Al_{0.62}N$, there is a change in the peak position of feature F along all the contact length. Higher Al-content TiAlN has a faster evolution of the spinodal decomposition, thus it is likely that TiN-enriched domains have formed in this coating also where the temperature has been lower [3].

The appearance of feature G, for $Ti_{0.38}Al_{0.62}N$, which is only observed for pure TiN in the as-deposited coatings, indicates that pure TiN is formed in the crater region of this coating. EXAFS reveals that in the middle of the crater, the average interatomic Ti-N distance has increased during use of the tool for both the $Ti_{0.47}Al_{0.53}N$ and $Ti_{0.38}Al_{0.62}N$ samples. The lattice parameter is expected to be larger in TiN-rich domains [2]. For the worn $Ti_{0.38}Al_{0.62}N$ sample, the Ti-N distance is similar to that of TiN, thus the TiN-rich domains that have formed have a composition close to pure TiN. For the $Ti_{0.47}Al_{0.53}N$ sample, the Ti-N interatomic distance is smaller for the worn sample compared to the worn $Ti_{0.38}Al_{0.62}N$, thus there is likely still some Al within the Ti-rich domains.

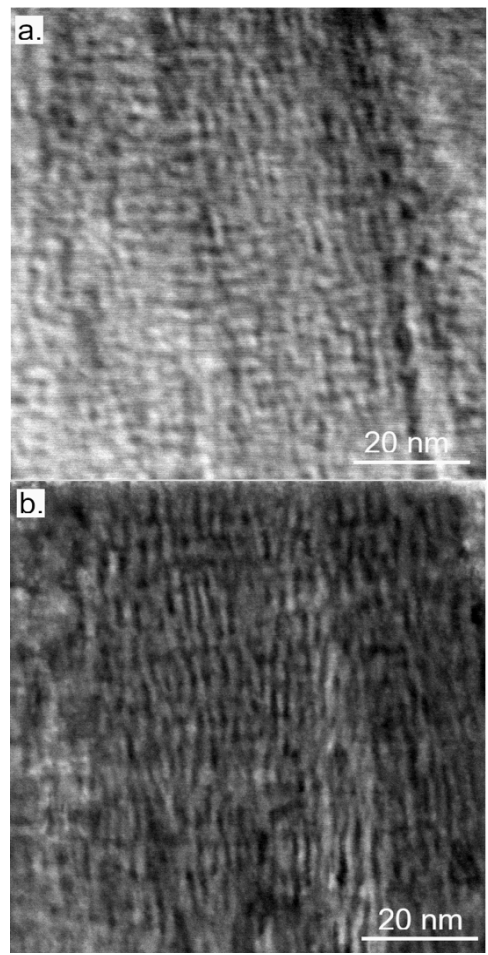

**Figure 7:** Elemental contrast STEM micrographs from the crater region of (a) $Ti_{0.47}Al_{0.53}N$ and (b) $Ti_{0.38}Al_{0.62}N$.

The XANES spectra suggest that there are TiN-enriched domains forming also in the fine-grained sample, thus spinodal decomposition occurs also in the fcc-phase of this coating. The shift of feature F is apparent across the entire contact (Fig. 3), thus spinodal decomposition occurs for all measured positions. A fine-grained microstructure has previously been observed to lower the temperature required for spinodal decomposition, and this might be the case also here, since the Al-content of the fine-grained fcc-phase is lower compared to the coarse-grained sample with similar overall Al-content [29].

Figure 8 shows the XANES spectra from the Cr 1$s$ and Fe 1$s$ absorption edges. The small SEM micrograph in Figure 8e illustrates the position where the spectra were recorded (compare with Figure 3). The white arrow illustrates the increasing distance from the tool edge and has a length of 500 µm. The EDS map of the worn inserts reveal that both Fe and Cr species from the stainless steel are found across the contact area (Fig. 3). For Cr, the spectra change appearance across the tool edge. For the first two positions, recorded in the crater region, the spectra are similar to that of the workpiece material. At the end of the crater region and the end of contact (yellow and blue), the spectra look different and are more similar to that of $Cr_2O_3$ [30, 31]. At the end of the contact length, there is access to oxygen from the surrounding air that can react with the tool and workpiece material, thus it is likely that Cr-O compounds form in this region. This has also been observed in our previous study [11]. This kind of compound can act as a protective layer during turning operation, reducing friction between coating and chip, and delaying crater formation [32, 33]. Even further away from the edge (purple and pink), the spectra are again similar to that of the steel workpiece. For Cr, the structure of the adhered material is like that of stainless steel in the crater region. At this position of the tool the temperature is expected to be low, thus reactions with oxygen from the air does not occur here. For the TiN coating, the signal is weak for the last two spectra indicating that only small amounts of Cr are present on the surface. For Fe, the spectra do not change across the contact length of the tool and is similar for both TiN and $Ti_{0.47}Al_{0.53}N$ coatings. The spectra are similar to the spectra recorded for the 316L stainless steel used here (dashed line) as well as that of a similar stainless steel (SS304) [34]. Thus, the Fe adhered to the tool has an fcc structure similar to that of the original stainless steel and no Fe-O phases form.

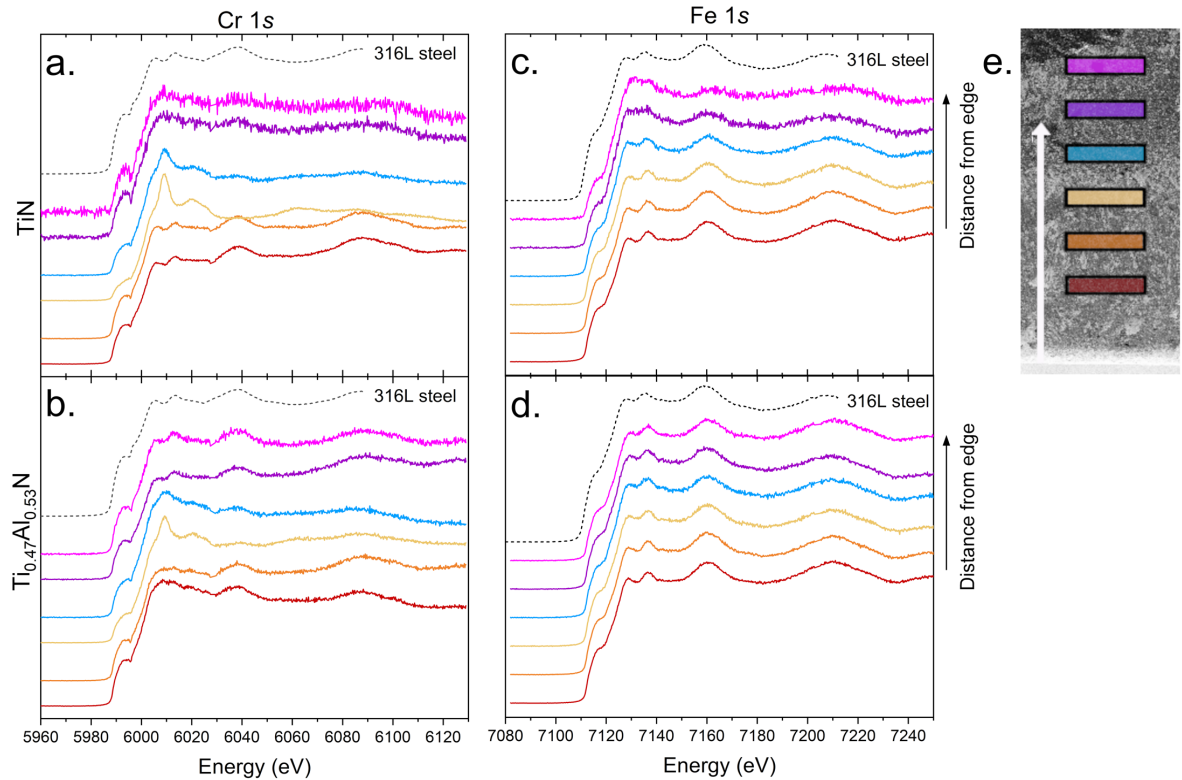

**Figure 8:** XANES data from the Cr 1*s* (a-b) and the Fe 1*s* edge (c-d) from (a,c) TiN and (b,d) Ti$_{0.47}$Al$_{0.53}$N coated inserts. The dashed line in (a-b) shows the data from the 316L steel reference sample. (e) illustrates the x-ray beam position for each scan (compare Fig. 3). The white arrow in (e) is 500 μm long.

## 4. Conclusion

Changes in atomic arrangements in the crater region of Ti$_{1-x}$Al$_x$N coated metal cutting tools during a turning operation were identified by XANES and confirmed by STEM. After 1 min of turning in 316L stainless steel, spinodal decomposition occurred in the Ti$_{0.47}$Al$_{0.53}$N and Ti$_{0.38}$Al$_{0.62}$N coatings. The XANES spectrum is sensitive to the Ti-content of the fcc-TiAlN phase and it enables identification of the formation of TiN-enriched domains during spinodal decomposition. A XANES linescan reveals that the largest amount of TiN-rich domains form at 300-350 μm from the edge of the tool. In the fine-grained Ti$_{0.38}$Al$_{0.62}$N sample, there is spinodal decomposition of the fcc-TiAlN as well as changes of the w-TiAlN phase during turning. For both coarse-grained and fine-grained Ti$_{0.38}$Al$_{0.62}$N, XANES reveals that there is a change in the local structure of Ti across all the contact length, suggesting that spinodal decomposition is initiated at the full length of the contact area after 1 min of turning.

Steel species adheres to the tool rake face during metal machining. Adhered Cr is oxidized in the end of the contact length where the temperature is high and there is access to O from the air. Adhered Fe remains in the fcc phase of the stainless steel and does not form Fe-O bonds.

The high sensitivity of XANES to the chemical composition of the fcc-TiAlN phase and the small beam size provided by the Balder beamline, enabling high spatial resolution, makes it a promising technique for identification of local phase changes in hard coatings.

**Acknowledgements**

We acknowledge the Swedish Governmental Agency for Innovation Systems (Vinnova grant no. 2018-04417) for financial support and the staff at MAX IV Laboratory for experimental support. This study was performed within the framework of the competence center FunMat-II that is financially supported by VINNOVA (grant no 2016-05156). We also acknowledge Janella Salamania for providing magnetron sputtered reference samples.

**References**

[1] R. Rachbauer, S. Massl, E. Stergar, D. Holec, D. Kiener, J. Keckes, J. Patscheider, M. Stiefel, H. Leitner, P.H. Mayrhofer, Decomposition pathways in age hardening of Ti-Al-N films, J. Appl. Phys. 110 (2011) 023515,https://doi.org/10.1063/1.3610451.
[2] L. Rogström, J. Ullbrand, J. Almer, L. Hultman, B. Jansson, M. Odén, Strain evolution during spinodal decomposition of TiAlN thin films, Thin Solid Films 520 (2012) 5542,https://doi.org/10.1016/j.tsf.2012.04.059.
[3] A. Knutsson, J. Ullbrand, L. Rogström, N. Norrby, L.J.S. Johnson, L. Hultman, J. Almer, M.P.J. Jöesaar, B. Jansson, M. Odén, Microstructure evolution during the isostructural decomposition of TiAlN - A combined in-situ small angle x-ray scattering and phase field study, J. Appl. Phys. 113 (2013) 213518,https://doi.org/10.1063/1.4809573.
[4] N. Norrby, L. Rogström, M.P. Johansson-Jõesaar, N. Schell, M. Odén, In situ X-ray scattering study of the cubic to hexagonal transformation of AlN in $Ti_{1-x}Al_xN$, Acta Mater. 73 (0) (2014) 205,http://dx.doi.org/10.1016/j.actamat.2014.04.014.
[5] K.M. Calamba, I.C. Schramm, M.P.J. Jõesaar, J. Ghanbaja, J.F. Pierson, F. Mücklich, M. Odén, Enhanced thermal stability and mechanical properties of nitrogen deficient titanium aluminum nitride ($Ti_{0.54}Al_{0.46}N_y$) thin films by tuning the applied negative bias voltage, J. Appl. Phys. 122 (6) (2017) 065301,https://doi.org/10.1063/1.4986350.
[6] A. Knutsson, M.P. Johansson, L. Karlsson, M. Odén, Thermally enhanced mechanical properties of arc evaporated $Ti_{0.34}Al_{0.66}N$/TiN multilayer coatings, J. Appl. Phys. 108 (4) (2010) 044312,https://doi.org/10.1063/1.3463422.
[7] M. Bartosik, H.J. Böhm, C. Krywka, Z.L. Zhang, P.H. Mayrhofer, Influence of phase transformation on the damage tolerance of Ti-Al-N coatings, Vacuum 155 (2018) 153,https://doi.org/10.1016/j.vacuum.2018.06.001.
[8] N. Norrby, M.P. Johansson, R. M'Saoubi, M. Odén, Pressure and temperature effects on the decomposition of arc evaporated $Ti_{0.6}Al_{0.4}N$ coatings in continuous turning, Surf. Coat. Technol. 209 (0) (2012) 203,http://dx.doi.org/10.1016/j.surfcoat.2012.08.068.
[9] M.P. Johansson Jõesaar, N. Norrby, J. Ullbrand, R. M'Saoubi, M. Odén, Anisotropy effects on microstructure and properties in decomposed arc evaporated $Ti_{1-x}Al_xN$ coatings during metal cutting, Surf. Coat. Technol. 235 (0) (2013) 181,http://dx.doi.org/10.1016/j.surfcoat.2013.07.031.
[10] K.M. Calamba, M.P. Johansson Jõesaar, S. Bruyère, J.F. Pierson, R. Boyd, J.M. Andersson, M. Odén, The effect of nitrogen vacancies on initial wear in arc deposited ($Ti_{0.52},Al_{0.48}$)$N_y$, (y < 1) coatings during machining, Surface and Coatings Technology 358 (2019) 452,https://doi.org/10.1016/j.surfcoat.2018.11.062.

[11] M. Moreno, J.M. Andersson, R. Boyd, M.P. Johansson-Jöesaar, L.J.S. Johnson, M. Odén, L. Rogström, Crater wear mechanism of TiAlN coatings during high-speed metal turning, Wear 484-485 (2021) 204016,https://doi.org/10.1016/j.wear.2021.204016.

[12] C. Wustefeld, D. Rafaja, M. Dopita, M. Motylenko, C. Baehtz, C. Michotte, M. Kathrein, Decomposition kinetics in Ti1-xAlxN coatings as studied by in-situ X-ray diffraction during annealing, Surf. Coat. Technol. 206 (7) (2011) 1727,https://doi.org/10.1016/j.surfcoat.2011.09.041.

[13] M.H. Tuilier, M.J. Pac, G. Covarel, C. Rousselot, L. Khouchaf, Structural investigation of thin films of Ti1−xAlxN ternary nitrides using Ti K-edge X-ray absorption fine structure, Surf. Coat. Technol. 201 (8) (2007) 4536,https://doi.org/10.1016/j.surfcoat.2006.09.095.

[14] M.-H. Tuilier, M.-J. Pac, M. Gîrleanu, G. Covarel, G. Arnold, P. Louis, C. Rousselot, A.-M. Flank, Electronic and atomic structures of Ti1−xAlxN thin films related to their damage behavior, J. Appl. Phys. 103 (8) (2008) 083524,https://doi.org/10.1063/1.2907415.

[15] M. Gîrleanu, M.J. Pac, O. Ersen, J. Werckmann, G. Arnold, C. Rousselot, M.H. Tuilier, The role of structural properties on damage behaviour of titanium and aluminium nitride coatings: An EXAFS and TEM study, Surf. Coat. Technol. 204 (12) (2010) 2042,https://doi.org/10.1016/j.surfcoat.2009.10.029.

[16] R. Gago, A. Redondo-Cubero, J.L. Endrino, I. Jiménez, N. Shevchenko, Aluminum incorporation in Ti1−xAlxN films studied by x-ray absorption near-edge structure, J. Appl. Phys. 105 (11) (2009) 113521,https://doi.org/10.1063/1.3139296.

[17] J.L. Endrino, C. Arhammar, A. Gutiérrez, R. Gago, D. Horwat, L. Soriano, G. Fox-Rabinovich, D. Martín y Marero, J. Guo, J.E. Rubensson, J. Andersson, Spectral evidence of spinodal decomposition, phase transformation and molecular nitrogen formation in supersaturated TiAlN films upon annealing, Acta Mater. 59 (16) (2011) 6287,https://doi.org/10.1016/j.actamat.2011.06.039.

[18] R. Gago, F. Soldera, R. Hübner, J. Lehmann, F. Munnik, L. Vázquez, A. Redondo-Cubero, J.L. Endrino, X-ray absorption near-edge structure of hexagonal ternary phases in sputter-deposited TiAlN films, J. All. Comp. 561 (2013) 87,https://doi.org/10.1016/j.jallcom.2013.01.130.

[19] K. Klementiev, K. Norén, S. Carlson, K.G.V.S. Clauss, I. Persson, The BALDER Beamline at the MAX IV Laboratory, J. Phys. Conf. Ser. 712 (2016) 012023,https://doi.org/10.1088/1742-6596/712/1/012023.

[20] J. Salamania, D.G. Sangiovanni, A. Kraych, K.M. Calamba, I.C. Schramm, L.J.S. Johnson, R. Boyd, B. Bakhit, T.W. Hsu, M. Mrovec, L. Rogström, F. Tasnádi, I.A. Abrikosov, M. Odén, Dislocation core structures in titanium nitride, Submitted for publication (2022)

[21] K.M. Calamba, J. Salamania, M.P.J. Jõesaar, L.J.S. Johnson, R. Boyd, J.F. Pierson, M.A. Sortica, D. Primetzhofer, M. Odén, Effect of nitrogen vacancies on the growth, dislocation structure, and decomposition of single crystal epitaxial (Ti1-xAlx)Ny thin films, Acta Mater. 203 (2021) 116509,https://doi.org/10.1016/j.actamat.2020.116509.

[22] K.V. Klementev, Extraction of the fine structure from x-ray absorption spectra, J. Phys. D: Appl. Phys. 34 (2) (2001) 209,https://doi.org/10.1088/0022-3727/34/2/309.

[23] J.J. Rehr, J.J. Kas, M.P. Prange, A.P. Sorini, Y. Takimoto, F. Vila, Ab initio theory and calculations of X-ray spectra, C. R. Phys. 10 (6) (2009) 548,https://doi.org/10.1016/j.crhy.2008.08.004.

[24] J.J. Rehr, J.J. Kas, F.D. Vila, M.P. Prange, K. Jorissen, Parameter-free calculations of X-ray spectra with FEFF9, Phys. Chem. Chem. Phys. 12 (21) (2010) 5503,https://doi.org/10.1039/B926434E.

[25] M. Magnuson, L.-Å. Näslund, Local chemical bonding and structural properties in Ti3AlC2 MAX phase and Ti3C2Tx MXene probed by Ti 1s x-ray absorption spectroscopy, Phys. Rev. Res. 2 (3) (2020) 033516,https://doi.org/10.1103/PhysRevResearch.2.033516.

[26] R.B. Greegor, F.W. Lytle, D.R. Sandstrom, J. Wong, P. Schultz, Investigation of TiO2   SiO2 glasses by X-ray absorption spectroscopy, J. Non-Cryst. Solids 55 (1) (1983) 27,https://doi.org/10.1016/0022-3093(83)90005-4.

[27] D. Cabaret, A. Bordage, A. Juhin, M. Arfaouia, E. Gaudryad, First-principles calculations of X-ray absorption spectra at the K-edge of 3d transition metals: an electronic structure analysis of the pre-edge, Phys. Chem. Chem. Phys. 12 (2010) 5619,https://doi.org/10.1039/B926499J.

[28] C.A. Triana, C.M. Araujo, R. Ahuja, G.A. Niklasson, T. Edvinsson, Electronic transitions induced by short-range structural order in amorphous ${\mathrm{TiO}}_{2}$, Phys. Rev. B 94 (16) (2016) 165129,https://doi.org/10.1103/PhysRevB.94.165129.

[29] A.B.B. Chaar, L. Rogström, M.P. Johansson-Jöesaar, J. Barrirero, H. Aboulfadl, N. Schell, D. Ostach, F. Mücklich, M. Odén, Microstructural influence of the thermal behavior of arc deposited TiAlN coatings with high aluminum content, J. All. Comp. 854 (2021) 157205,https://doi.org/10.1016/j.jallcom.2020.157205.

[30] M.L. Peterson, G.E. Brown, G.A. Parks, C.L. Stein, Differential redox and sorption of Cr(III/IV) on natural silicate and oxide minerals: EXAFS and XANES results, Geochim. Cosmochim. Acta 61 (16) (1997) 3399,https://doi.org/10.1016/S0016-7037(97)00165-8.

[31] S. Bell, M.W.M. Jones, E. Graham, D.J. Peterson, G.A.v. Riessen, G. Hinsley, T. Steinberg, G. Will, Corrosion mechanism of SS316L exposed to NaCl/Na2CO3 molten salt in air and argon environments, Corros. Sci. 195 (2022) 109966

[32] P.V. Moghaddam, B. Prakash, E. Vuorinen, M. Fallqvist, J.M. Andersson, J. Hardell, High temperature tribology of TiAlN PVD coating sliding against 316L stainless steel and carbide-free bainitic steel, Tribology International 159 (2021) 106847,https://doi.org/10.1016/j.triboint.2020.106847.

[33] X.D. Fang, D. Zhang, An investigation of adhering layer formation during tool wear progression in turning of free-cutting stainless steel, Wear 197 (1) (1996) 169,https://doi.org/10.1016/0043-1648(96)06924-4.

[34] D. Garai, A. Wilson, I. Carlomagno, C. Meneghini, F. Carla, H. Hussain, A. Gupta, J. Zegenhagen, Structure of the Surface Region of Stainless Steel: Bulk and Thin Films, Phys. Status Solidi B 259 (2022) 2100513, https://doi.org/10.1002/pssb.202100513.